# Analytical characterization of self-sustained nonlinear oscillators modelling human walking and bouncing

Varun Nevash[1], Prakash Kumar[2*], Chinika Dangi[3]

[1]Department of Mechanical Engineering and Applied Mechanics, University of Pennsylvania, Philadelphia, 19104, PA, USA

[2]Department of Production Engineering, National Institute of Technology, Tiruchirappalli, Tiruchirappalli, 620015, India

[3]Department of Mechanical Engineering, Indian Institute of Technology Jodhpur, N.H. 62, Nagaur Road, Karwar, Jodhpur, 342030, India

*Correspondence: **prakashkumar@nitt.edu**

**Abstract**

Researchers have developed hybrid Van der Pol–Rayleigh–Duffing–type oscillators to model human-induced forces; however, their analytical framework has largely relied on the Lindstedt–Poincare perturbation method, energy balance approaches, and harmonic balance techniques. This paper aims to apply new mathematical tools to these existing models and address potential research gaps. An analytical proof for the stability of the limit cycle has been formulated by using the Krylov–Bogolyubov perturbation method. The multiple-scales method has been modified to highlight an iterative algorithm for determining the order of approximation required to capture nonlinear effects. The describing function method is utilised to formulate an alternate amplitude. Comparisons between first-order amplitudes obtained from perturbation analysis and the describing function formulations reveal conditions under which the two approaches converge. These conditions are exploited to formulate additional constraints for the estimation of model parameters, offering a systematic alternative to purely optimisation-based approaches.

**Keywords:** Self-sustained nonlinear oscillators, Multiple-scales method, Describing function method, Model parameters, Perturbation methods, Amplitude.

## 1. Introduction

The recent advancements in Computer-Aided Engineering and data-driven decision systems have revolutionised the field of structural design, leading to the development of high-strength materials and advanced construction techniques to build aesthetically appealing and high-performing slender structures. These light structures, owing to characteristics of low stiffness and damping, possess low natural frequencies often below 5Hz, susceptible to human-induced excitations leading to resonance [1]. Structures such as pedestrian bridges, buildings, and floors are prone to the crowd action of human rhythmic activities, including walking, running, jumping, and bouncing, which can result in a modification of the designed structural response and compromise safety. This has led to the closure of various structures like the Millennium Bridge in London, the T-Bridge in Japan, and the Solferino Bridge in Paris, spurring significant research in the study of crowd-structure interactions [2,3].

Silvano et al. [4] have classified the existing literature on crowd-structure interaction for walking into five classes based on coupling interactions between pedestrians and pedestrians, and between pedestrians and structures, characteristics of the floor (oscillating or rigid), and the phenomenon of movement (crowd or single pedestrian). In class A, both coupling interactions are considered for crowd walking on a moving floor, whereas in class B, coupling with structure is not considered and walking is on a rigid floor. In class C, only pedestrian-structure interaction is considered for single/few pedestrians walking on a moving floor. In classes D and E, the uncoupled problem with a single pedestrian walking on an oscillating floor and a rigid floor is considered, respectively. This classification can be extended and generalised for other human rhythmic activities. In this paper, we focus on Class E (Uncoupled problem with a single pedestrian bouncing on a rigid floor).

Several experimental analyses and theoretical models have been developed for single pedestrian movement on a rigid floor. These can be classified broadly into three categories, i.e., Statistical, Stochastic, and Mechanical system-based approaches [5]. The statistical approach is based on replicating the frequency content of the human rhythmic activity by using Ground Reaction Force (GRF). This force is shown to be periodic and therefore can be modelled using a truncated Fourier series with a Dynamic load factor (DLF) (amplitudes of harmonic components in series normalized by body weight) [6]. The major limitation of this method is that it does not account for structural vibrations in the estimation of pedestrian forces; it cannot capture high inter- and intra-subject variability [7]. The stochastic modelling approach overcomes the previous limitation, but doesn't capture higher-order dynamical instability [8]. Physics-based modelling approaches involving mechanical systems have been effectively able to capture human rhythmic activities. Biomechanically inspired models involving a family of inverted pendulum models, like Spring Loaded Inverted Pendulum (SLIP) and Damped Bipedal Inverted Pendulum (DBIP), have also been studied. The major drawback is their inability to accurately reproduce higher-order DLFs [9,10].

Human rhythmic activities are self-sustained, as they internally produce energy to sustain motion. Consequently, they are modelled using self-sustained oscillators with a stable limit cycle; those shapes and sizes are close to the periodic orbits in the phase-plane diagram obtained from experimental data [4]. Researchers have employed hybrid models of Vander-Pol, Rayleigh, and Duffing oscillators to describe bouncing and walking, i.e., lateral, longitudinal, and vertical forces independently [4,5,11–15]. The key advantage of the proposed modelling approach is that pedestrian–structure interaction can be easily integrated, as the Ground Reaction Force (GRF) generated by the oscillators is a function of the displacement, velocity, and acceleration of the body with respect to the structure [15]. A detailed draft of assumptions used, data collection, and modelling approach can be found in [15]. For the benefit of readers, we have presented a brief mathematical overview.

The dynamics of the pedestrian can be modelled as a Single Degree of Freedom mechanical system.

$$m\ddot{u}(t) + F(t) = 0 \qquad (1.1)$$

where *m* denotes the mass of pedestrian, assumed to be lumped at the centre of gravity (CoG), and $\ddot{u}(t)$ represents the acceleration of CoG. The term F(t) represents the GRF.

Eq. (1.1) represents an oscillator that is assumed to have the following dynamic characteristics:

1. No external excitations (no time-varying input acting on the system) [16].

2. Autonomous (driven by self-governed dynamics, usually represented by Ordinary Differential Equations (ODE) that do not explicitly depend on time) [16].

3. The solution $u(t)$ represents the oscillations of the Centre of gravity of the human about a straight trajectory parallel to the axis of the force being modelled.

The variables $\ddot{u}(t), \dot{u}(t), u(t)$, represent the acceleration, velocity, and displacement of the CoG, respectively, and $F(t)$ is the human force deduced from experimental data. The acceleration can be directly obtained from Eq. (1.1), and velocity and displacement can be obtained from the time-domain integration of Eq. (1.1). The motion described by $u(t)$ under these assumptions provides a close approximation of the CoG motion of a person.

A self-sustained oscillator is a system that generates and maintains stable oscillations without an external force input [17]. The equation of motion of a harmonic oscillator can be represented as Eq. (1.2),

$$\ddot{u} + 2\mu\omega_0\,\dot{u}\bigl(1 - h(u,\dot{u})\bigr) + \omega_0^2 u = 0 \qquad (1.2)$$

where $\omega_0 > 0$ is the circular frequency of the underlying linear system, and $\mu$ is strictly positive. We also suppose that $h$ is a polynomial of the variables $u$ and $\dot{u}$. An exception to this form (1.2) of harmonic oscillators is (1.3), where the damping coefficient has an acceleration term. To the best of the authors' knowledge, these models cover the entirety of the literature in self-sustained oscillators developed for modelling human rhythmic activities.

Silvano et al. [15] developed the first model for lateral walking force (1.5), later two models for lateral (1.3) and (1.4), one each for vertical (1.6) and longitudinal (1.7) walking forces were developed by Kumar et al. [11–14]. Recently, a model was made for the bouncing force [5].

$$\ddot{u} - 2\mu\omega_0\dot{u}\left(1 - \beta u^2 - \frac{\delta}{\omega_0^2}\dot{u}^2 - \frac{\psi}{\omega_0^3}\ddot{u}\dot{u}\right) + 2\mu\omega_0^2\kappa u^3 + \omega_0^2 u = 0 \qquad (1.3)$$

$$\ddot{u} - 2\mu\omega_0\dot{u}\left(1 - \beta u^2 - \frac{\delta}{\omega_0^2}\dot{u}^2 - \frac{\xi}{\omega_0^3}\dot{u}^3 u\right) + 2\mu\omega_0^2\lambda u^5 + \omega_0^2 u = 0 \qquad (1.4)$$

$$\ddot{u} - 2\mu\omega_0\dot{u}\left(1 - \beta u^2 - \frac{\gamma}{\omega_0}\dot{u}u - \frac{\delta}{\omega_0^2}\dot{u}^2\right) + \omega_0^2 u = 0 \qquad (1.5)$$

$$\ddot{u} - 2\mu\omega_0\dot{u}\left(1 - \frac{\eta}{\omega_0}\dot{u} - \beta u^2 - \frac{\delta}{\omega_0^2}\dot{u}^2 - \frac{\vartheta}{\omega_0}\dot{u}u^2\right) + 2\mu\omega_0^2 u^3(\kappa + u^2\lambda) + \omega_0^2 u = 0 \qquad (1.6)$$

$$\ddot{u} - 2\mu\omega_0\dot{u}\left(1 - \frac{\eta}{\omega_0}\dot{u} - vu - \frac{\delta}{\omega_0^2}\dot{u}^2\right) + \omega_0^2 u = 0 \qquad (1.7)$$

$$\ddot{u} - 2\mu\omega_0\dot{u}\left(1 - \frac{\eta}{\omega_0}\dot{u} - \frac{\lambda}{\omega_0}\dot{u}u - \beta u^2 - \frac{\delta}{\omega_0^2}\dot{u}^2\right) + 2\mu\omega_0^2\kappa u^2 + \omega_0^2 u = 0 \qquad (1.8)$$

In the above equations, $\mu$ is the nonlinearity associated with the model, and the model parameters $\psi, \eta, \delta, \vartheta, \beta,$ and $\kappa, \lambda$ are associated with nonlinear damping and stiffness terms, respectively.

An acceleration term in the damping coefficient might clutter the mathematical framework developed in this paper, which is applicable to the vast majority of oscillators. So, we shall focus on five existing models (1.4 – 1.8) developed for describing human walking and bouncing [4,5,12–14]. For brevity, mostly the results have been presented only for the oscillator modelling vertical human walking force [14]. Oscillators modelling lateral [4,12,13] human walking consist only of odd power terms, whereas the oscillators modelling vertical [14] walking force consist of both odd and even power terms along with two nonlinear stiffness terms; moreover, it has the highest number of model parameters, making it mathematically the most complicated oscillator in this literature.

The analytical results presented for these models are only based on the Lindstedt-Poincare perturbation method and the energy balance method. In this article, we apply new mathematical frameworks to analyse these oscillators. Throughout this paper, the numerical data used for simulations are taken directly from [4,5,12–14]. In Section 2, the linearised behaviour of the system near equilibrium is examined by formulating the Jacobian matrix. The next part provides the foundational transformations used in subsequent sections. In Section 4, we apply the Krylov-Bogolyubov perturbation method to investigate the stability of the limit cycle. We highlight potential limitations of the Lindstedt-Poincare perturbation method and discuss interesting inferences that could be obtained through the multiple-scales method in Section 5. Section 6 discusses how the describing function method can be used to formulate amplitudes, and when the results from this method match the ones from perturbation analysis, and for weak non-linearities, how this could be exploited to improve the estimation of model parameters. Finally, Sections 7 and 8 present future insights and conclusions, respectively.

2. **Linearized behavior of phase-portrait near the origin: Jacobian Matrix**

For a non-zero value of $\mu$, the Hartman-Grobman theorem [18] enables us to study the primary behavior of the phase portrait near the equilibrium point by using the Jacobian matrix [19]. The oscillator modelling vertical walking forces (2.1) is studied across sections 2 – 5.

$$\ddot{u} - 2\mu\omega_0 \dot{u}\left(1 - \frac{\eta}{\omega_0}\dot{u} - \beta u^2 - \frac{\delta}{\omega_0^2}\dot{u}^2 - \frac{\vartheta}{\omega_0}\dot{u}u^2\right) + 2\mu\omega_0^2 u^3(\kappa + u^2\lambda) + \omega_0^2 u = 0 \quad (2.1)$$

Let $u = x, \dot{u} = y$ & $A$ be the Jacobian matrix of the system. Then, (2.2) expresses a linearization of the system about the equilibrium point.

$$\begin{bmatrix}\dot{x}\\\dot{y}\end{bmatrix} = A\begin{bmatrix}x\\y\end{bmatrix} \quad (2.2)$$

$$A = \begin{bmatrix}\frac{\partial \dot{x}}{\partial x} & \frac{\partial \dot{x}}{\partial y}\\ \frac{\partial \dot{y}}{\partial x} & \frac{\partial \dot{y}}{\partial y}\end{bmatrix} = \begin{bmatrix}0 & 1\\ \frac{\partial \dot{y}}{\partial x} & \frac{\partial \dot{y}}{\partial y}\end{bmatrix} \quad (2.3)$$

$$= \begin{bmatrix}0 & 1\\ 2\mu\omega_0 y\left(-2\beta x - \frac{2\vartheta}{\omega_0}yx\right) - 2\mu\omega_0^2 x^3(3\kappa + 5x^2\lambda) - \omega_0^2 & 2\mu\omega_0\left(1 - \frac{2\eta}{\omega_0}y - \beta x^2 - \frac{3\delta}{\omega_0^2}(y)^2 - \frac{2\vartheta}{\omega_0}yx^2\right)\end{bmatrix} \quad (2.4)$$

The origin is the only equilibrium point for these systems. At origin, all these models (1.4 – 1.8) have the same Jacobian matrix (2.5).

$$A = \begin{bmatrix} 0 & 1 \\ -\omega_0^2 & 2\mu\omega_0 \end{bmatrix} \qquad (2.5)$$

For weak non-linearity $(0 < \mu < 1)$, we obtain $\Delta = \text{Determinant}(A) = \omega_0^2 >, \tau = \text{trace}(A) = 2\mu\omega_0 > 0$ and $\tau^2 - 4\Delta < 0$. From the characteristics of a Jacobian matrix [18], we can conclude that these systems have an unstable spiral near the origin. This can also be seen in Figure 1.

3. **Overview of amplitude-phase dynamics: Transformations and phase portrait.**

Transformations involving timescales and coordinates catalyze our understanding of a complex system. Transforming the equation (2.1) into a different timescale does not affect the existence and stability of the limit cycle. It helps us in eliminating the frequency (ω) parameter, thus making our equation handy for applications involving complex computations.

Let $\varepsilon = 2\mu$ and $\tau = \omega t$ then $u(\tau) = u(t), \omega \dot{u} = u'$ and $\omega^2 \ddot{u} = u''$ where $\dot{u}, \ddot{u}$ and $u', u''$ represent first-order and second-order differentiation with respect to $t$ and $\tau$ respectively. Transforming (2.1), we get (3.1),

$$u'' - \varepsilon u'(1 - \eta u' - \beta u^2 - \delta(u')^2 - \vartheta u' u^2) + \varepsilon u^3(\kappa + u^2\lambda) + u = 0 \qquad (3.1)$$

By Van der Pol transformation [19] on (3.1), we obtain (3.2),

$$u' = v$$

$$v' = \varepsilon u'(1 - \eta u' - \beta u^2 - \delta(u')^2 - \vartheta u' u^2) - \varepsilon u^3(\kappa + u^2\lambda) - u \qquad (3.2)$$

We can see that the only equilibrium points are $(u, v) = (0,0)$. So, if a periodic solution exists, it must enclose the origin. The Van der Pol transformation converts the second order (3.1) into a system of first order differential equations (3.2), which is predominantly used in numerical simulations.

Transforming a system into polar coordinates helps us to visualize the model in trigonometric forms, enabling more characterization. This formalization of the equation would be particularly useful when we have to develop an analytical proof for the existence of a limit cycle [18]. Without loss of generality, on the phase portrait, we can assume that $(u, v) = (r\cos\theta, r\sin\theta)(r \in (0, +\infty) \ \& \ \theta \in [0,2\pi))$. We have $u' = r'\cos\theta - r\theta'\sin\theta \ \& \ v' = r'\sin\theta + r\theta'\cos\theta$ where $r'$ and $\theta'$ represent the first-order differentiation of $r$ and $\theta$ with respect to $\tau$ respectively.

By substituting the polar coordinates in (3.3),

$$v' = \varepsilon r \sin\theta (1 - \eta r \sin\theta - \beta r^2 \cos^2\theta - \delta r^2 \sin^2\theta - \vartheta r^3 \cos^2\theta \sin\theta)$$
$$- \varepsilon r^3 \cos^3\theta (\kappa + \lambda r^2 \cos^2\theta) - r\cos\theta$$

$$u' = r\sin\theta \tag{3.3}$$

From equations (3.3) and the expression for $u'$ & $v'$, we get (3.4),

$$r' = -\varepsilon r \sin^2\theta (\eta r \sin\theta + \beta r^2 \cos^2\theta + \delta r^2 \sin^2\theta + \vartheta r^3 \cos^2\theta \sin\theta - 1)$$
$$- \varepsilon r^3 \cos^3\theta \sin\theta (\kappa + \lambda r^2 \cos^2\theta)$$

$$\theta' = \frac{r'}{r\tan\theta} - 1 \tag{3.4}$$

Kumar et al. [14] have presented the limit cycle of a particular pedestrian F at a walking speed of 4.5 km/h. In Figure 1, we have extended this to all walking speeds for the same pedestrian F. Here, this exercise is carried out solely for the sake of completeness.

## 4. Stability of limit cycle: Krylov–Bogolyubov perturbation method

Results from the characterization of the Jacobian matrix fail to capture the dynamics of the phase portrait far away from the equilibrium point due to the linearization of a system involving higher-order nonlinear terms. Perturbation methods [20] helps us to understand the asymptotic behavior of a nonlinear system.

Krylov–Bogolyubov perturbation method [21] is used to study the stability of oscillators with weak non-linearities. $\mu \ll 1$. Like other perturbation methods, in addition to estimating the first-order amplitude and frequency, it also provides inference about the stability of the limit cycle.

$$\ddot{u} - 2\mu\omega_0 \dot{u}\left(1 - \frac{\eta}{\omega_0}\dot{u} - \beta u^2 - \frac{\delta}{\omega_0^2}\dot{u}^2 - \frac{\vartheta}{\omega_0}\dot{u}u^2\right) + 2\mu\omega_0^2 u^3(\kappa + u^2\lambda) + \omega_0^2 u = 0 \tag{4.1}$$

In (4.1), introducing state variables $u = x$ & $\dot{u} = y$. Then $\dot{y} = -\varepsilon h(x, y) - \omega_0^2 x$, where,

$$h = -\omega_0 y\left(1 - \frac{\eta}{\omega_0}y - \beta x^2 - \frac{\delta}{\omega_0^2}y^2 - \frac{\vartheta}{\omega_0}yx^2\right) + 2\mu\omega_0^2 x^3(\kappa + x^2\lambda) \tag{4.2}$$

Let $x(t) = r\cos(s)$ then $y = -r\omega \sin(s)$, where $s = \phi(t) + \omega t$.

$$(x\omega)^2 + y^2 = (r\omega)^2$$

$$r\dot{r} = x\dot{x} + y\dot{y}$$

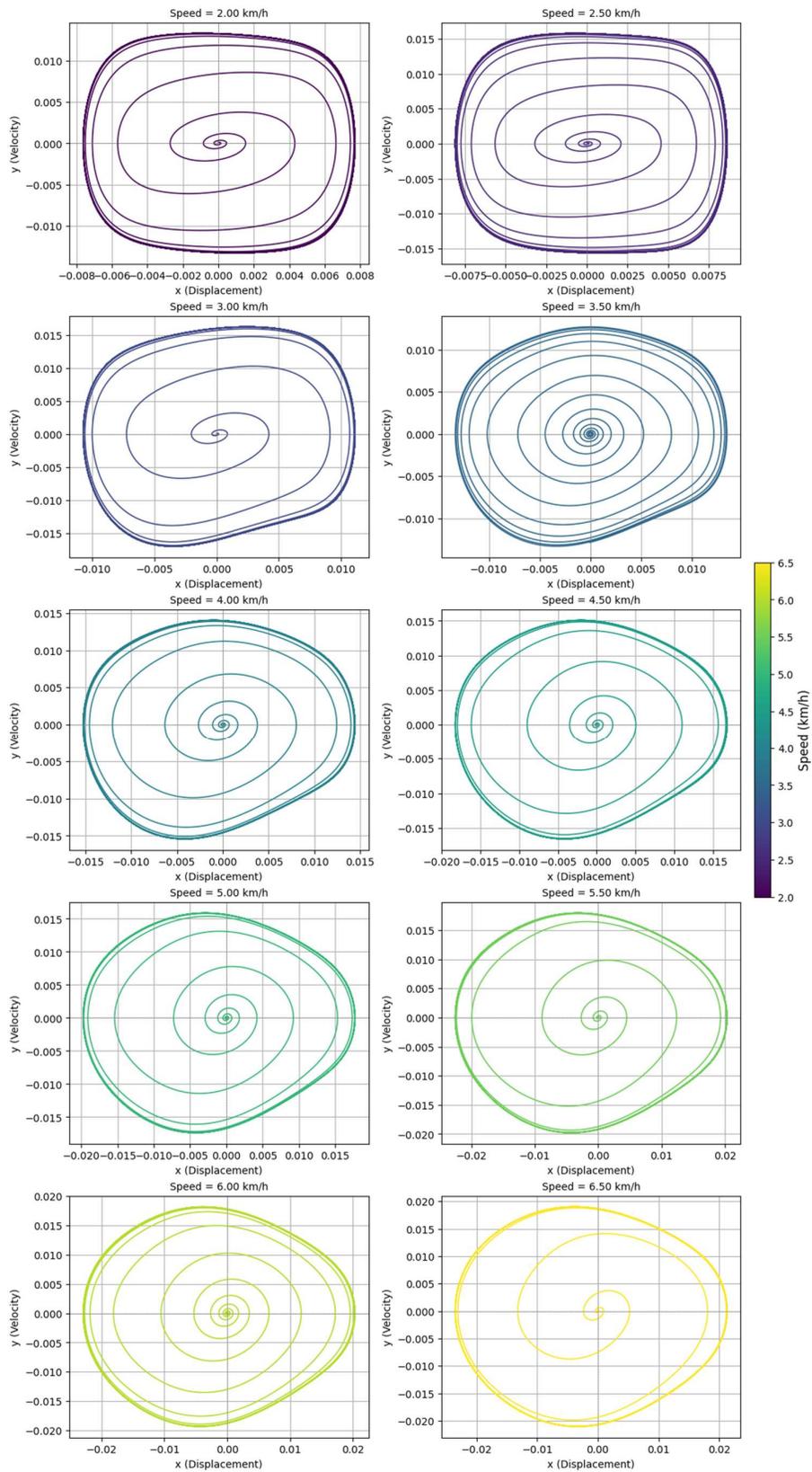

**Figure 1: Phase portrait of oscillators modelling vertical walking force.**

$$\tan(s) = -\frac{y}{x\omega} \tag{4.3}$$

On simplification, we get (4.4),

$$\dot{r} = \frac{(1-\omega_0^2)xy - \varepsilon hy}{r} \tag{4.4}$$

On averaging,

$$\overline{hy} = \left< -\omega_0 y^2 \left(1 - \frac{\eta}{\omega_0}y - \beta x^2 - \frac{\delta}{\omega_0^2}y^2 - \frac{\vartheta}{\omega_0}yx^2\right) + 2\mu\omega_0^2 x^3 y(\kappa + x^2\lambda) \right>$$

$$= \left< -\omega_0{}^3 r^2 \sin^2 s(1 + \eta r \sin s - \beta r^2 \cos^2 s - \delta r^2 \sin^2 s + \vartheta r^3 \cos^2 s \sin s) - \right.$$
$$\left. -\omega_0{}^3 \varepsilon r^4 \sin s \cos^3 s \,(\kappa + \lambda r^2 \cos^2 s) \right>$$

$$= -\omega_0{}^3 \bar{r}^2 < (\sin^2 s(1 + \eta r \sin s - \beta r^2 \cos^2 s - \delta r^2 \sin^2 s + \vartheta r^3 \cos^2 s \sin s) -$$
$$\varepsilon r^2 \sin s \cos^3 s \,(\kappa + \lambda r^2 \cos^2 s) >$$

$$\tag{4.5}$$

As $<\sin^n\theta\cos^m\theta> = 0$ if m or n is odd and $<\sin^2\theta\cos^2\theta> = \frac{1}{8}, <\sin^2\theta> = \frac{1}{2}, <\sin^4\theta> = \frac{3}{8}$. Then, $\dot{\bar{r}} = -\varepsilon\omega_0{}^3\bar{r}(\frac{1}{2} - \frac{3}{8}\delta\bar{r}^2 - \frac{1}{8}\beta\bar{r}^2)$, and $\dot{\bar{r}} = 0$ when $\bar{r} = 0$ or $\bar{r} = \frac{2}{\sqrt{\beta+3\delta}}$.

$$\dot{\phi} = \frac{\varepsilon h \cos(s)}{r\omega}$$

$$\dot{\bar{\phi}} = \frac{\varepsilon < h\cos(s) >}{r\omega} \tag{4.6}$$

As, $<\cos^4\theta> = \frac{3}{8}, <\sin^6\theta> = \frac{5}{16}$. Then, $\dot{\bar{\phi}} = \varepsilon\omega_0\bar{r}^2\left(\frac{6\kappa+5\lambda\bar{r}^2}{16}\right)$.

A stable limit cycle is a closed trajectory $C$ in phase space that is a solution of a nonlinear differential equation and has the property that all neighbouring trajectories in the phase plane spiral towards the curve $C$ as time $t$ tends to infinity; if instead they spiral away from it, the limit cycle is termed unstable or non-attractive. [19].

By substituting the first-order amplitude $\bar{r}$ expression, we get $\frac{d\dot{\bar{r}}}{dr} = \varepsilon\omega_0{}^3(\frac{1}{2} - \frac{9}{8}\delta\bar{r}^2 - \frac{3}{8}\beta\bar{r}^2) = -1$. As, $\frac{d\dot{\bar{r}}}{dr}$ always negative for $\bar{r} = \frac{2}{\sqrt{\beta+3\delta}}$, so it is a stable limit cycle, and $\frac{d\dot{\bar{\phi}}}{d\omega} = \mu(\frac{3\kappa(\beta+3\delta)+10\lambda}{(\beta+3\delta)^2})$. Thus, the system has an unstable spiral leading to a stable limit cycle. This can also be seen in Figure 1.

## 5. Iterative multiple-scales method for time-scale identification

In the Lindstedt-Poincare perturbation method, we find an approximate analytical solution by assuming periodic behavior using a frequency-induced time scale [4]. In the multiple-scales method [20] we analyze the system at different time scales. For the solution of the system, approximated up to $O(\varepsilon^k)$ and valid up to the time scale $O(\varepsilon^{k-n})$, we need to consider n time scales. This approach could serve as a yardstick to determine the number of time scales we need to introduce to achieve a determined accuracy in the approximation.

$$U(t, \varepsilon t, \varepsilon^2 t, \varepsilon^3 t, \varepsilon^4 t, \ldots, \varepsilon^n t) = U_0(t, \varepsilon t, \varepsilon^2 t, \varepsilon^3 t, \varepsilon^4 t, \ldots, \varepsilon^n t) +$$
$$\varepsilon U_1(t, \varepsilon t, \varepsilon^2 t, \varepsilon^3 t, \varepsilon^4 t, \ldots, \varepsilon^n t) + \varepsilon^2 U_2(t, \varepsilon t, \varepsilon^2 t, \varepsilon^3 t, \varepsilon^4 t, \ldots, \varepsilon^n t) + \cdots \quad (5.1)$$

$$U_0(t, \varepsilon t, \varepsilon^2 t, \varepsilon^3 t, \varepsilon^4 t, \ldots, \varepsilon^n t) = A_0(\varepsilon t, \varepsilon^2 t, \varepsilon^3 t, \varepsilon^4 t, \ldots, \varepsilon^n t) \cos(t +$$
$$\theta_0(\varepsilon t, \varepsilon^2 t, \varepsilon^3 t, \varepsilon^4 t, \ldots, \varepsilon^n t)) \quad (5.2)$$

$$\forall\, j \geq 1,\ U_j(t, \varepsilon t, \varepsilon^2 t, \varepsilon^3 t, \varepsilon^4 t, \ldots, \varepsilon^n t) = A_j(\varepsilon t, \varepsilon^2 t, \varepsilon^3 t, \varepsilon^4 t, \ldots, \varepsilon^n t) \cos\Big(t +$$
$$\theta_j(\varepsilon t, \varepsilon^2 t, \varepsilon^3 t, \varepsilon^4 t, \ldots, \varepsilon^n t)\Big) + \sum_{m \in \mathbb{N}} C_m \sin(m\theta) + D_m \cos(m\theta) \quad (5.3)$$

We can find $C_m$ & $D_m$ the coefficients of higher order harmonics using the particular integral of the equation associated with $O(\varepsilon^j)$, likewise $A_j$ & $\theta_j$ can be found by using the initial conditions, and the eradication of secular terms in the equation associated with $O(\varepsilon^{j+1})$.

Consider the general equation representing oscillators,

$$u'' + u = \varepsilon f(u, u'), \quad \varepsilon \ll 1 \quad (5.4)$$

Let $T = \varepsilon t$ and $U(t, T, \varepsilon) = u(t, \varepsilon)$

$$\frac{du}{dt} = \frac{\partial U}{\partial t} + \varepsilon \frac{\partial U}{\partial T}, \quad \frac{d^2 u}{dt^2} = \frac{\partial^2 U}{\partial t^2} + 2\varepsilon \frac{\partial^2 U}{\partial t\, \partial T} + \varepsilon^2 \frac{\partial^2 U}{\partial T^2} \quad (5.5)$$

Substituting (5.5) in (5.4),

$$\frac{\partial^2 U}{\partial t^2} + 2\varepsilon \frac{\partial^2 U}{\partial t\, \partial T} + \varepsilon^2 \frac{\partial^2 U}{\partial T^2} + U = \varepsilon f\left(U, \frac{\partial U}{\partial t} + \varepsilon \frac{\partial U}{\partial T}\right) \quad (5.6)$$

For our oscillator modelling vertical walking force,

$$f(u, u') = \omega_0 u'\left(1 - \frac{\eta}{\omega_0} u' - \beta u^2 - \frac{\delta}{\omega_0^2}(u')^2 - \frac{\vartheta}{\omega_0} u'u^2\right) - \omega_0^2 u^3(\kappa + u^2 \lambda) \quad (5.7)$$

Now using an asymptotic power series expansion of $U$ (5.8) in (5.6) yields (5.9).

$$U(t,T,\varepsilon) \simeq \sum_{j=0}^{N} \varepsilon^j U_j(t,T) + o(\varepsilon^N) \tag{5.8}$$

$$\sum_{j=0}^{N} \varepsilon^j \frac{\partial^2 U_j}{\partial t^2} + 2\varepsilon^{j+1} \frac{\partial^2 U_j}{\partial t\, \partial T} + \varepsilon^{j+2} \frac{\partial^2 U_j}{\partial T^2} + \varepsilon^j U_j = \varepsilon f \left( \sum_{j=0}^{N} \varepsilon^j U_j, \sum_{j=0}^{N} \varepsilon^j \frac{\partial U_j}{\partial t} + \varepsilon^{j+1} \frac{\partial U_j}{\partial T} \right)$$

$$= \varepsilon f\left(U_0, \frac{\partial U_0}{\partial t}\right) + \varepsilon^2 U_1 f_u\left(U_0, \frac{\partial U_0}{\partial t}\right) + \varepsilon^2 \left(\frac{\partial U_1}{\partial t} + \frac{\partial U_0}{\partial T}\right) f_{u'}\left(U_0, \frac{\partial U_0}{\partial t}\right) + \mathcal{O}(\varepsilon^3) \tag{5.9}$$

where $f_u$, $f_{u'}$ denote first-order partial differentiation with respect to $u$ & $u'$ respectively.

Suppose the initial conditions are given by (5.10),

$$\begin{cases} u(0) = a_0 \\ u'(0) = 0 \end{cases} \tag{5.10}$$

Then for $U_j$ the new initial conditions are (5.11) and (5.12),

$$U_0(0,0) = a, \quad \frac{\partial U_0}{\partial t}(0,0) = 0 \tag{5.11}$$

$$U_j(0,0) = 0, \quad \frac{\partial U_j}{\partial t}(0,0) = -\frac{\partial U_{j-1}}{\partial T}(0,0) \quad \forall j \geq 1 \tag{5.12}$$

Solving for leading order terms: $O(\varepsilon^0)$

$$\frac{\partial^2 U_0}{\partial t^2} + U_0 = 0 \tag{5.13}$$

The general solution is given by $U_0(t,T) = A_0(T)\cos(t + \theta_0(T))$. The initial conditions then lead to

$$\begin{cases} A_0(0)\cos(\theta_0(0)) = a_0 \\ -A_0(0)\sin(\theta_0(0)) = 0 \end{cases} \tag{5.14}$$

This implies that for $A_0 \neq 0$, we have $A_0(0) = a_0$, $\theta_0(0) = 0$.

Substituting $U_0(t,T) = A_0(T)\cos(\phi), \phi = t + \theta_0(T)$ into $f(U_0, U_0')$ we get,

$$f(U_0, U_0') = -\omega_0^2 A_0 \sin \phi \, (1 + \eta A_0 \sin \phi - \beta A_0^2 \cos^2 \phi - \delta A_0^2 \sin^2 \phi$$

$$+ \vartheta A_0^3 \cos^2 \phi \sin \phi) - \omega_0^2 A_0^3 \cos^3 \phi \, (\kappa + \lambda A_0^2 \cos^2 \phi) \tag{5.15}$$

Using the Fourier expansion [20], we can derive the solvability conditions (5.16) and (5.17),

$$2\frac{dA_0}{dT} + b_{11}(T) = 0 \tag{5.16}$$

$$2A_0\frac{d\theta_0}{dT} + a_{11}(T) = 0 \tag{5.17}$$

, and the Fourier coefficients (5.18 – 5.20).

$$a_{10} = \frac{1}{2\pi}\int_0^{2\pi} f\left(U_0(t,T), \frac{\partial U_0(t,T)}{\partial t}\right) dt \tag{5.18}$$

$$a_{1n} = \frac{1}{\pi}\int_0^{2\pi} f\left(U_0(t,T), \frac{\partial U_0(t,T)}{\partial t}\right) \cos(n\phi) dt \tag{5.19}$$

$$b_{1n} = \frac{1}{\pi}\int_0^{2\pi} f\left(U_0(t,T), \frac{\partial U_0(t,T)}{\partial t}\right) \sin(n\phi) dt \tag{5.20}$$

Solving for amplitude in (5.21 – 5.23),

$$\frac{dA_0}{dT} = \frac{-1}{2\pi}\int_0^{2\pi} (-\omega_0^2 A_0 \sin\phi \, (1 + \eta A_0 \sin\phi - \beta A_0^2\cos^2\phi - \delta A_0^2\sin^2\phi$$
$$+ \vartheta A_0^3\cos^2\phi \sin\phi) - \omega_0^2 A_0^3\cos^3\phi \, (\kappa + \lambda A_0^2\cos^2\phi)) \sin(\phi_0) dt$$

$$\frac{dA_0}{dT} = \frac{A_0}{2}\left(1 - \frac{(\beta + 3\delta)}{4}A_0^2\right) \tag{5.21}$$

$$A_0 = \frac{\frac{1}{\sqrt{C}}}{\sqrt{\frac{\frac{1}{C} - a_0^2}{a_0^2}e^{-T} + 1}} \tag{5.22}$$

$$C = \frac{(\beta + 3\delta)}{4} \tag{5.23}$$

Similarly, by solving for $\theta_0$ in (5.24), we get $\theta_0(T) = \theta_0(0) = 0$ & $U_0 = \frac{\frac{1}{\sqrt{C}}}{\sqrt{\frac{\frac{1}{C} - a_0^2}{a_0^2}e^{-T} + 1}}\cos t$.

$$\frac{d\theta_0}{dT} = \frac{-1}{2\pi}\int_0^{2\pi} -\omega_0^2 A_0 \sin\phi \, (1 + \eta A_0 \sin\phi - \beta A_0^2\cos^2\phi - \delta A_0^2\sin^2\phi$$
$$+ \vartheta A_0^3\cos^2\phi \sin\phi) - \omega_0^2 A_0^3\cos^3\phi \, (\kappa + \lambda A_0^2\cos^2\phi) \cos(\phi_0) dt = 0$$

$$\tag{5.24}$$

We can see that $A_0(T)$ is bounded and approaches $\frac{2}{\sqrt{\beta+3\delta}}$ as $T \to \infty$. We need to compute higher-order terms to understand the dynamics arising due to non-linearities.

$$\frac{\partial^2 U_1}{\partial t^2} + U_1 = -2\frac{\partial^2 U_0}{\partial t \partial T} + f\left(U_0, \frac{\partial U_0}{\partial t}\right) \tag{5.25}$$

$$\frac{\partial^2 U_1}{\partial t^2} + U_1 + 2\frac{\partial^2 U_0}{\partial t \partial T} - \frac{\partial U_0}{\partial t} + \eta(\frac{\partial U_0}{\partial t})^2 + \beta\frac{\partial U_0}{\partial t}U_0^2 + \delta(\frac{\partial U_0}{\partial t})^3 + \vartheta(\frac{\partial U_0}{\partial t})^2 U_0^2 + \kappa U_0^3 + \lambda U_0^5 = 0 \tag{5.26}$$

Solving by substituting results of (5.24), we get (5.27)

$$U_1 = A_1(T)\cos(t + \theta_1(T)) - \left(\frac{\eta A_0^2(T)}{2} + \frac{\vartheta A_0^4(T)}{8}\right) - \frac{(\beta-\delta)A_0^3(T)}{32}\sin(3t) - \frac{\eta A_0^2(T)}{6}\cos(2t) +$$
$$\left(\frac{\kappa A_0^3(T)}{32} + \frac{5\lambda A_0^5(T)}{128}\right)\cos(3t) - \frac{\vartheta A_0^4(T)}{120}\cos(4t) + \frac{\lambda A_0^5(T)}{384}\cos(5t) \tag{5.27}$$

Applying initial conditions (5.11) and (5.12),

$U_1(0,0) = 0$:

$$A_1(0)\cos(\theta_1(0)) = \frac{2\eta A_0^2(0)}{3} + \frac{2\vartheta A_0^4(0)}{15} - \frac{\kappa A_0^3(0)}{32} - \frac{\lambda A_0^5(0)}{24} \tag{5.28}$$

as $A_0(0) = a_0$,

$$A_1(0)\cos(\theta_1(0)) = \frac{2\eta a_0^2}{3} + \frac{2\vartheta a_0^4}{15} - \frac{\kappa a_0^3}{32} - \frac{\lambda a_0^5}{24} \tag{5.29}$$

$\frac{\partial U_1}{\partial t}(0,0) + \frac{\partial U_0}{\partial T}(0,0) = 0$:

$$A_1(0)\sin(\theta_1(0)) = -\frac{(\beta-\delta)a_0^3}{32} + \frac{(1-Ca_0^2)a_0}{2} \tag{5.30}$$

To find $A_1(T)$, we need to formulate $O(\varepsilon^2)$ terms.

$$\frac{\partial^2 U_2}{\partial t^2} + 2\frac{\partial^2 U_1}{\partial t \partial T} + \frac{\partial^2 U_0}{\partial T^2} - \frac{\partial U_1}{\partial t} - \frac{\partial U_0}{\partial T} + 2\eta\frac{\partial U_0}{\partial t}\left(\frac{\partial U_1}{\partial t} + \frac{\partial U_0}{\partial T}\right) + \beta\left(2U_0\frac{\partial U_0}{\partial t}\frac{\partial U_1}{\partial t} + \frac{\partial U_1}{\partial t}U_0^2 +\right.$$
$$\left.\frac{\partial U_0}{\partial T}U_0^2\right) + \delta(\frac{\partial U_0}{\partial t})^2\left(3\frac{\partial U_1}{\partial t} + 3\frac{\partial U_0}{\partial T}\right) + \vartheta(U_0^2(\frac{\partial U_1}{\partial t})^2\frac{\partial U_0}{\partial t} + 2\left((\frac{\partial U_0}{\partial t})^2 U_0 U_1 + 2\frac{\partial U_0}{\partial T}U_0^2\frac{\partial U_0}{\partial t} +\right.$$
$$\left.U_0^2\frac{\partial U_0}{\partial t}\frac{\partial U_1}{\partial t}\right) + 3\kappa U_0^2 U_1 + 5\lambda U_0^4 U_1 + U_2 = 0 \tag{5.31}$$

So, as to completely compute $U_1(t, T)$, we can use the expressions derived using the initial conditions (5.28 – 5.30) and equate the coefficients of secular terms in equation (5.31) associated with $O(\varepsilon^2)$ to zero.

Segregating the secular terms in (5.31),

$-2A_1' \sin(t + \theta_1) - 2A_1 \cos(t + \theta_1) \theta_1' + A_1 \sin(t + \theta_1) - A_0' \cos t + A_0'' \cos t +$

$\beta(\frac{A_0^2 A_1}{2} \cos(t - \theta_1) + \frac{3(\beta-\delta)A_0^5}{64} \sin t - \frac{A_0^2 A_1}{2} \sin(t + \theta_1) + \frac{A_0^2 A_1}{4} \sin(t - \theta_1) + \frac{3A_0^2 A_0'}{4} \cos t +$

$(\frac{3\kappa A_0^5}{64} + \frac{15\lambda A_0^7}{256}) \cos t) + 5\lambda(\frac{3A_0^4 A_1}{8} \cos(t + \theta_1) + \frac{A_0^4 A_1}{4} \cos(t - \theta_1) - \frac{3(\beta-\delta)A_0^7}{512} \sin t + (\frac{5\kappa A_0^7}{512} +$

$\frac{77\lambda A_0^9}{6144}) \cos t) + \delta(\frac{3A_0^2 A_0'}{4} \cos t - \frac{3A_0^2 A_1}{2} \sin(t + \theta_1) + \frac{3A_0^2 A_1}{4} \sin(t - \theta_1) + \frac{9(\beta-\delta)A_0^5}{128} \cos t +$

$\frac{9\kappa A_0^5}{128} \sin t) + 3\kappa(\frac{A_0^2 A_1}{2} \cos(t - \theta_1) + \frac{A_0^2 A_1}{2} \cos(t + \theta_1) - \frac{(\beta-\delta)A_0^5}{128} \sin t + (\frac{\kappa A_0^5}{128} + \frac{5\lambda A_0^7}{512}) \cos t) -$

$\frac{A_0^3 \eta^2}{3} \cos t - \vartheta(\sin t (\frac{A_0^2 A_1^2}{8} + \frac{A_0^6 \eta^2}{72} + \frac{9(\beta-\delta)^2 A_0^8}{8192} + \frac{9\kappa^2 A_0^8}{8192} + \frac{225\lambda^2 A_0^{12}}{131072} + \frac{45\lambda\kappa A_0^{10}}{8192}) + (\frac{A_0^7 \vartheta}{32} + \frac{A_0^5 \eta}{8} +$

$\frac{A_1 A_0^4 \eta}{12} \sin\theta_1) \cos t)$  (5.32)

Equating coefficients of $\sin t$ in (5.32) to zero,

$-2\frac{dA_1}{dT} \cos\theta_1 + 2\frac{d\theta_1}{dT} A_1 \sin\theta_1 + A_1 \cos\theta_1 - \frac{\beta A_0^2 A_1}{4} \cos\theta_1 + \frac{3\beta(\beta-\delta)A_0^5}{64} +$

$\frac{\beta A_0^2 A_1}{2} \sin\theta_1 + 5\lambda(-\frac{3(\beta-\delta)A_0^7}{512} - \frac{A_0^4 A_1}{8} \sin\theta_1) + \delta(\frac{9\kappa A_0^5}{128} - \frac{3A_0^2 A_1}{4} \cos\theta_1) - \frac{3\kappa(\beta-\delta)A_0^5}{128} -$

$\vartheta(\frac{A_0^2 A_1^2}{8} + \frac{A_0^6 \eta^2}{72} + \frac{9(\beta-\delta)^2 A_0^8}{8192} + \frac{9\kappa^2 A_0^8}{8192} + \frac{225\lambda^2 A_0^{12}}{131072} + \frac{45\lambda\kappa A_0^{10}}{8192}) = 0$  (5.33)

Equating coefficients of $\cos t$ in (5.32) to zero,

$-2\frac{dA_1}{dT} \sin\theta_1 - 2A_1 \cos\theta_1 \frac{d\theta_1}{dT} + A_1 \sin\theta_1 - \frac{dA_0}{dT} + A_0'' - \frac{3\beta A_0^2 A_1}{4} \sin\theta_1 + \frac{3\beta A_0^2 A_0'}{4} + \beta(\frac{3\kappa A_0^5}{64} +$

$\frac{15\lambda A_0^7}{256}) + \frac{\beta A_0^2 A_1}{2} \cos\theta_1 + 5\lambda((\frac{5\kappa A_0^7}{512} + \frac{77\lambda A_0^9}{6144}) + \frac{5A_0^4 A_1}{8} \cos\theta_1) + \delta(\frac{3A_0^2 A_0'}{4} + \frac{9(\beta-\delta)A_0^5}{128} -$

$\frac{3A_0^2 A_1}{4} \sin\theta_1) + 3\kappa((\frac{\kappa A_0^5}{128} + \frac{5\lambda A_0^7}{512}) + A_0^2 A_1 - \frac{A_0^3 \eta^2}{3} - \vartheta(\frac{A_0^7 \vartheta}{32} + \frac{A_0^5 \eta}{8} + \frac{A_1 A_0^4 \eta}{12} \sin\theta_1) = 0$  (5.34)

Remember in (5.32 – 5.34), $A_0, A_0', A_0''$ & $C$ represent $A_0 = \dfrac{\frac{1}{\sqrt{C}}}{\sqrt{\frac{\frac{1}{C}-a_0^2}{a_0^2}e^{-T}+1}}$, $A_0' =$

$\dfrac{(\frac{\frac{1}{C}-a_0^2}{a_0^2}e^{-T})}{2\sqrt{C}(\frac{\frac{1}{C}-a_0^2}{a_0^2}e^{-T}+1)^{3/2}}$, $A_0'' = \dfrac{-(\frac{\frac{1}{C}-a_0^2}{a_0^2}e^{-T})(\frac{\frac{5}{C}-a_0^2}{2\,a_0^2}e^{-T}+1)}{2\sqrt{C}(\frac{\frac{1}{C}-a_0^2}{a_0^2}e^{-T}+1)^{5/2}}$ & $C = \dfrac{(\beta+3\delta)}{4}$ respectively.

We have two differential equations (5.33, 5.34) in two variables ($A_1, \theta_1$). We realize that these coupled differential equations are quite complicated to solve analytically. Nevertheless, the purpose of this exercise was to demonstrate the application of the multiple-scales method for understanding the extent of approximation we must make while solving complex systems numerically.

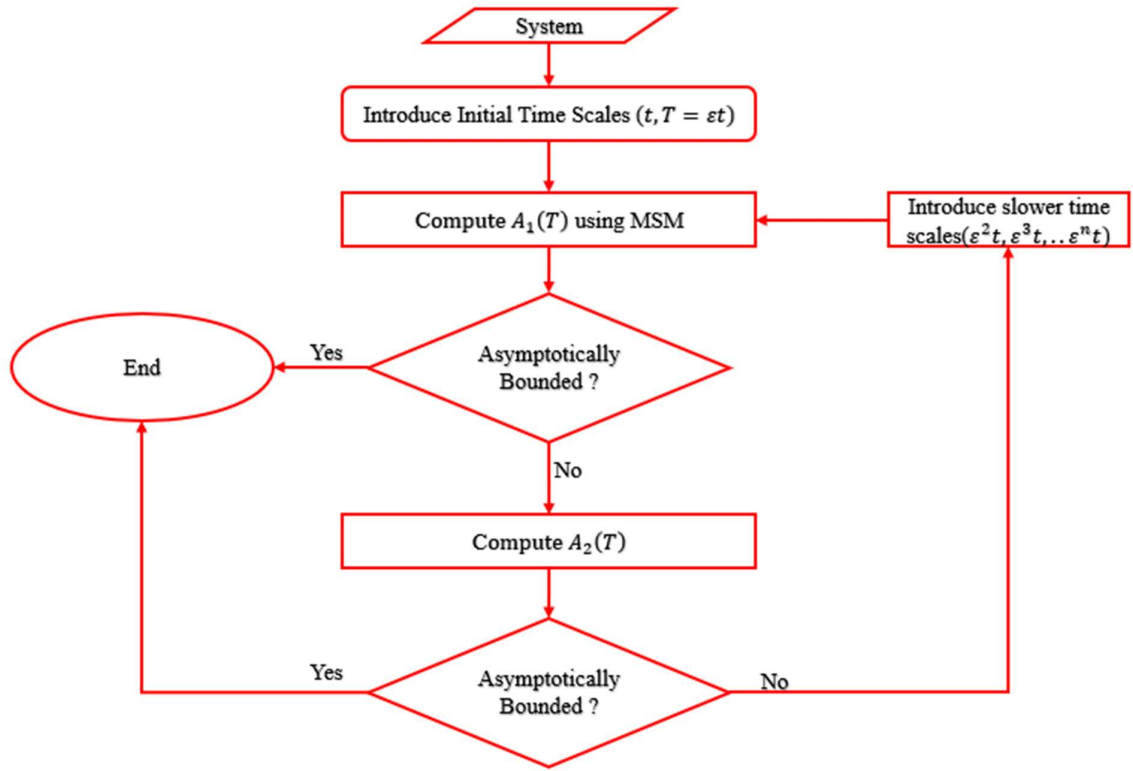

**Figure 2: Flowchart illustrating the procedure for identifying the appropriate timescale and order of estimation for dynamical systems.**

If the solution $A_1(T)$ has a bounded asymptotic limit, then for the chosen system with time scale valid up to $O(\varepsilon^{-1})$, the approximation captures all nonlinear effects. In case, $A_1(T)$ does not show bounded behavior, which mostly occurs due to the presence of positive powers of $T$, then we may have to iterate this procedure and check for $A_2(T), A_3(T), \ldots A_b(T)$ until we get a bounded amplitude, there is also a possibility that we might have no bounded solution obtained by considering just two time scales. So, we may want to analyze the phenomenon at slower time scales, i.e., take $t, T = \varepsilon t, \tau = \varepsilon^2 t$, repeat the entire analytical calculation done in this section and check for bounded solutions. These iterations would give us a sense of the timescale at which the system approaches a limit cycle. Thus, the multiple-scales method can be used to understand the rate of convergence of amplitude over different scales of time.

The flowchart presented in Figure 2 gives an outline of this iterative procedure. For most research applications, with a given timescale, we check for a bounded solution up to second-order amplitude; if it doesn't exist, we dilute the timescale to achieve bounded behavior.

6. **Formulation of the relationship between Model parameters: Describing function method**

Describing function method [22,23] is an excellent tool for analyzing non-linear oscillators exhibiting periodic behavior. From equation (6.1), we can extract the linear and non-linear elements.

$$\ddot{u} - 2\mu\omega_0\dot{u} + \omega_0^2 u + 2\mu\omega_0\dot{u}\left(\frac{\eta}{\omega_0}\dot{u} + \beta u^2 + \frac{\delta}{\omega_0^2}\dot{u}^2 + \frac{\vartheta}{\omega_0}\dot{u}u^2\right) + 2\mu\omega_0^2 u^3(\kappa + u^2\lambda) = 0 \quad (6.1)$$

**Linear element**

$$\ddot{u} - 2\mu\omega_0\dot{u} + \omega_0^2 u \quad (6.2)$$

Applying the Laplace transform, we get, $p^2 x(p) - 2\mu\omega_0 p x(p) + \omega_0^2 x(p) = y(p)$ where $p$ is the Laplace transform parameter, $x$ & $y$ correspond to transforms of input and output in the linear block.

$$G(p) = \frac{x(p)}{y(P)} \quad (6.3)$$

$$= \frac{1}{p^2 - 2\mu\omega_0 p + \omega_0^2} \quad (6.4)$$

**Non-linear element**

$$2\mu\omega_0\dot{u}\left(\frac{\eta}{\omega_0}\dot{u} + \beta u^2 + \frac{\delta}{\omega_0^2}(\dot{u})^2 + \frac{\vartheta}{\omega_0}\dot{u}u^2\right) + 2\mu\omega_0^2 u^3(\kappa + u^2\lambda) \quad (6.5)$$

Substituting $u = A\cos\omega t$, and $\dot{u} = -A\omega\sin\omega t$ in (6.5). We obtain (6.6) and (6.7), where $N_i$ and $N_o$ are input and output non-linear functions.

$$N_i = A\cos\omega t \quad (6.6)$$

$$N_o = -2\mu A\omega\omega_0\sin\omega t\left(-\frac{\eta}{\omega_0}A\omega\sin\omega t + \beta A^2\cos^2\omega t + \frac{\delta}{\omega_0^2}A^2\omega^2\sin^2\omega t -$$

$$\frac{\vartheta}{\omega_0}A^3\cos^2\omega t\sin\omega t\right) - 2\mu A^3\omega_0^2\cos^3\omega t\,(\kappa + \lambda A^2\cos^2\omega t) \quad (6.7)$$

Expressing (6.7) as a sum of harmonic multiples of cosines and sines, and neglecting higher order terms, we get $N_0 = 2\mu\omega_0\left(\frac{-\beta A^3\omega}{4} - \frac{3\delta A^3}{4w_0^2}\omega^3\right)\sin\omega t + 2\mu\omega_0^2\left(\frac{3\kappa A^3}{4} + \frac{5\lambda A^5}{8}\right)\cos\omega t$. Let $C_1 = \left(\frac{-\beta A^3\omega}{4} - \frac{3\delta A^3}{4w_0^2}\omega^3\right)$, and $C_2 = \omega_0\left(\frac{3\kappa A^3}{4} + \frac{5\lambda A^5}{8}\right)$, then $N_0 = 2\mu\omega_0 R\cos(\omega t - \phi)$ where

$\phi = tan^{-1}(C_1/C_2)$ and $R = \sqrt{C_1^2 + C_2^2}$. Consequently, the non-linear response function can be expressed as (6.8).

$$N(A,\omega) = \frac{N_0}{N_i} = \frac{2\mu\omega_0 R}{A}e^{-j\phi} \qquad (6.8)$$

From real and imaginary parts of (6.9), we get (6.10).

$$1 + G(j\omega)N(A,\omega) = 0 \qquad (6.9)$$

$$2\mu\omega\omega_0 = -\frac{2\mu\omega_0 R}{A}sin\phi$$

$$\omega^2 - \omega_0^2 = \frac{2\mu\omega_0 R}{A}cos\phi \qquad (6.10)$$

From (6.10), we have developed equations (6.11) for $A$ and $\omega$.

$$15A^6\delta\mu\lambda + 18A^4\kappa\delta\mu + 4((3\delta + \beta)A^2 - 16 = 0$$

$$\omega^2 = \frac{\omega_0^2(4 - \beta A^2)}{3\delta A^2} \qquad (6.11)$$

**Comparing numerical solutions of amplitude from the describing function method and the perturbation method.**

In Tables 2-5, the last two columns, $A_1$ and $A_2$, represent the first-order amplitude computed using perturbation analysis and describing function methods, respectively.

From Table 1, when $\mu \to 0$ or coefficients of higher-order powers of $\dot{u}$ tends to zero, i.e., in a cubic lateral and vertical oscillator when $\delta \to 0$, and in the quintic lateral oscillator when $\delta, \xi \to 0$, the characteristic polynomials obtained through the describing function method give $A = \frac{2}{\sqrt{\beta+3\delta}}$, similar to the first-order amplitude obtained by perturbation analysis. In vertical and cubic lateral oscillators at speeds of 5.25km/h and 3.5km/h, respectively, the value of $\delta$ estimated through the least square identification procedure is zero. As shown in Tables 2 and 3, we can observe that numerically the amplitudes formulated by the two methods are equal at these instances.

The sum of non-linear damping and stiffness terms, when expressed as a sum of harmonics, if it does not have any $cos\omega t$ term, then the first-order amplitude obtained through the perturbation analysis is the same as the one obtained through the describing function method. In the longitudinal oscillator, the non-linear damping term, when expressed as a sum of harmonics, has no $cos\omega t$ term, so we get the same expression for amplitude by both methods.

We observe that in the case of the bouncing model, the characteristic equation corresponding to amplitude gives us an alternate amplitude, i.e., a different mathematical expression which is numerically close to the first-order amplitude computed using perturbation analysis for all cases of model parameter estimated. Though the cubic lateral oscillator has few non-linear terms, higher symmetry and the same characteristic equation as the bouncing oscillator, the results obtained through the two methods do not converge. This contrast is because in a bouncing oscillator, the value of $\mu$ at all speeds is very small compared to the other four oscillators. Even in cubic lateral and vertical oscillators, when $\delta = 0$, the value of $\mu$ is very small compared to other speeds, the two methods are seen to converge in this instance. Thus, for oscillators with very weak non-linearity, the amplitude from perturbation analysis and the describing function method converge.

This observation can be used to formulate constraints for model parameters by equating the amplitude from the two methods. This technique can be useful in handling model estimation problems involving a system described by an array of differential equations similar to oscillators; additional constraints are known to improve the accuracy and computational time of optimisation problems [24].

For the bouncing model, $A^4 \beta\mu\lambda - A^2(6\delta + 2\beta + 4\mu\lambda) + 8 = 0$. The solution of the equation is, $A^2 = \frac{(6\delta+2\beta+4\mu\lambda) - \sqrt{(6\delta+2\beta+4\mu\lambda)^2 - 3}}{2\beta\mu\lambda}$. This is numerically close to the amplitude $A_0 = \frac{2}{\sqrt{3\delta+\beta}}$, as depicted in Table 4.

$$\frac{2}{\sqrt{3\delta + \beta}} = \sqrt{\frac{(6\delta + 2\beta + 4\mu\lambda) - \sqrt{(6\delta + 2\beta + 4\mu\lambda)^2 - 32\beta\mu\lambda}}{2\beta\mu\lambda}}$$

Figures 3-6 graphically depict the first-order amplitudes of oscillators obtained using the two methods. We can see that both methods show the same trend and tend to converge in instances of small $\mu$. This can be clearly seen in Figure 5, corresponding to the quintic lateral oscillator, as the responses never tend to converge at any walking speed due to considerable nonlinearity, even though $\mu < 1$. The describing function method incorporates a greater degree of nonlinearity in its first-order amplitude, making it more accurate than perturbation methods when dealing with considerable nonlinearity in the class of oscillators with weak nonlinearity.

**Table 1 – Characteristic equation for amplitude and frequency corresponding to different models.**

| Oscillator | Model | Amplitude Characteristic Equation | Frequency Characteristic Equation |
|---|---|---|---|
| Cubic Lateral | $\ddot{u} - 2\mu\omega_0\dot{u}\left(1 - \beta u^2 - \dfrac{\gamma}{\omega_0}\dot{u}u - \dfrac{\delta}{\omega_0^2}\dot{u}^2\right) + \omega_0^2 u = 0$ | $A^4\beta\mu\gamma - A^2(6\delta + 2\beta + 4\mu\gamma) + 8 = 0$ | $\omega^2 = \dfrac{\omega_0^2(4 - \beta A^2)}{3\delta A^2}$ |
| Bouncing | $\ddot{u} - 2\mu\omega_0\dot{u}\left(1 - \dfrac{\eta}{\omega_0}\dot{u} - \dfrac{\lambda}{\omega_0}\dot{u}u - \beta u^2 - \dfrac{\delta}{\omega_0^2}(\dot{u})^2\right)$ $+ 2\mu\omega_0^2\kappa u^2 + \omega_0^2 u = 0$ | $A^4\beta\mu\lambda - A^2(6\delta + 2\beta + 4\mu\lambda) + 8 = 0$ | $\omega^2 = \dfrac{\omega_0^2(4 - \beta A^2)}{3\delta A^2}$ |
| Vertical | $\ddot{u} - 2\mu\omega_0\dot{u}\left(1 - \dfrac{\eta}{\omega_0}\dot{u} - \beta u^2 - \dfrac{\delta}{\omega_0^2}\dot{u}^2 - \dfrac{\vartheta}{\omega_0}\dot{u}u^2\right)$ $+ 2\mu\omega_0^2 u^3(\kappa + u^2\lambda) + \omega_0^2 u = 0$ | $15A^6\delta\mu\lambda + 18A^4\kappa\delta\mu + 4(3\delta+\beta)A^2 - 16 = 0$ | $\omega^2 = \dfrac{\omega_0^2(4 - \beta A^2)}{3\delta A^2}$ |
| Quintic Lateral | $\ddot{u} - 2\mu\omega_0\dot{u}\left(1 - \beta u^2 - \dfrac{\delta}{\omega_0^2}\dot{u}^2 - \dfrac{\xi}{\omega_0^3}\dot{u}^3 u\right)$ $+ 2\mu\omega_0^2\lambda u^5 + \omega_0^2 u = 0$ | $(45\mu\lambda\delta^2 + \xi\mu\beta^2)A^6 - 8\beta\xi\mu A^4$ $+ (-36\delta^2 + 16\xi\mu - 12\delta\beta)A^2 + 48\delta$ $= 0$ | $\omega^2 = \dfrac{\omega_0^2(4 - \beta A^2)}{3\delta A^2}$ |
| Longitudinal | $\ddot{u} - 2\mu\omega_0\dot{u}\left(1 - \dfrac{\eta}{\omega_0}\dot{u} - \nu u - \dfrac{\delta}{\omega_0^2}\dot{u}^2\right) + \omega_0^2 u = 0$ | $A = 2/\sqrt{3\delta}$ | $\omega = \omega_0$ |

Table 2: Model parameters for vertical oscillator.

| Speed(km/h) | $\omega_0(rad/s)$ | $\mu$ | $\eta(m^{-1})$ | $\beta\ (10^3 m^{-2})$ | $\kappa(10^3 m^{-2})$ | $\delta(10^3 m^{-2})$ | $\vartheta\ (10^5 m^{-3})$ | $\lambda\ (10^6 m^{-4})$ | $A_1(m)$ | $A_2(m)$ |
|---|---|---|---|---|---|---|---|---|---|---|
| 2.0 | 4.24 | 0.188 | 0.0 | 0.0 | 0.0 | 6.730 | −3.76 | 4636.80 | 0.01408 | 0.01408 |
| 2.5 | 4.11 | 0.147 | 0.0 | 0.0 | 0.0 | 4.750 | −22.64 | 5400.81 | 0.01675 | 0.01675 |
| 3.0 | 5.83 | 0.259 | 17.05 | 21.22 | 0.0 | 1.480 | −9.15 | 451.20 | 0.02469 | 0.02469 |
| 3.5 | 10.01 | 0.077 | 78.82 | 21.16 | −49.77 | 0 | −21.85 | 358.90 | 0.04348 | 0.04344 |
| 4.0 | 10.98 | 0.136 | 58.39 | 11.54 | −18.02 | 2.230 | −9.35 | 100.31 | 0.02258 | 0.02259 |
| 4.5 | 12.82 | 0.147 | 52.46 | 6.18 | −14.26 | 2.940 | −5.67 | 41.91 | 0.02059 | 0.02059 |
| 5.0 | 13.88 | 0.140 | 56.18 | 4.64 | −12.94 | 3.060 | −5.08 | 30.32 | 0.02037 | 0.02037 |
| 5.5 | 14.64 | 0.184 | 35.86 | 3.05 | −7.30 | 2.600 | −2.36 | 12.81 | 0.02222 | 0.02222 |
| 6.0 | 15.65 | 0.108 | 60.56 | 2.91 | −12.54 | 2.680 | −4.05 | 20.65 | 0.02191 | 0.02191 |
| 6.5 | 15.87 | 0.269 | 23.10 | 2.13 | −3.83 | 2.570 | −1.27 | 5.98 | 0.02247 | 0.02247 |

Table 3: Model parameters for cubic lateral oscillator.

| Speed (km/h) | $\omega_0(rad/s)$ | $\mu$ | $\beta\ (m^{-2})$ | $\gamma(m^{-2})$ | $\delta(m^{-2})$ | $A_1(m)$ | $A_2(m)$ |
|---|---|---|---|---|---|---|---|
| 3.75 | 4.622 | 0.554 | 4784.6 | 7415.9 | 3426.0 | 0.01630 | 0.01365 |
| 4.50 | 4.943 | 0.621 | 6304.6 | 7452.0 | 3229.1 | 0.01582 | 0.01328 |
| 5.25 | 5.213 | 0.045 | 18278.80 | 103730.6 | 0 | 0.01479 | 0.01479 |
| 6.00 | 5.400 | 0.545 | 13964.9 | 12861.1 | 3117.5 | 0.01310 | 0.01135 |

Table 4: Model parameters for bouncing oscillator.

| Beat frequency(Hz) | $\omega_0(rad/s)$ | μ | β $(m^{-2})$ | δ$(m^{-2})$ | λ $(m^{-2})$ | η$(m^{-1})$ | κ$(m^{-1})$ | $A_1$(m) | $A_2$(m) |
|---|---|---|---|---|---|---|---|---|---|
| 1.5 | 8.99 | 0.054 | 621.79 | 0 | 1983.24 | 84.75 | 71.86 | 0.06389 | 0.06389 |
| 1.9 | 11.39 | 0.040 | 1102.34 | 0 | 2639.50 | 92.69 | 83.88 | 0.04701 | 0.04701 |
| 2.3 | 13.94 | 0.023 | 2421.61 | 0 | 4612.35 | 144.18 | 136.94 | 0.04008 | 0.04008 |
| 2.7 | 16.09 | 0.021 | 1029.00 | 999.98 | 5240.78 | 319.86 | 212.13 | 0.02860 | 0.02821 |
| 3.1 | 19.09 | 0.023 | 1525.61 | 1691.13 | 5346.48 | 503.40 | 304.16 | 0.01740 | 0.01732 |
| 3.5 | 20.79 | 0.023 | 2510.22 | 2039.35 | 5476.43 | 723.43 | 414.89 | 0.01286 | 0.01281 |

Table 5: Model parameters for quintic lateral oscillator.

| Speed(km/h) | $\omega_0(rad/s)$ | μ | β$(10^3 m^{-2})$ | δ$(10^3 m^{-2})$ | λ$(10^8 m^{-4})$ | ξ $(10^8 m^{-4})$ | $A_1$(m) | $A_2$(m) |
|---|---|---|---|---|---|---|---|---|
| 3.75 | 7.507 | 0.125 | 0 | 19.344 | −3.25 | 1.41 | 0.007713 | 0.008302 |
| 4.50 | 7.641 | 0.321 | 10.042 | 13.298 | −1.43 | 3.89 | 0.008594 | 0.008950 |
| 5.25 | 7.910 | 0.332 | 17.977 | 9.559 | −1.37 | 3.13 | 0.008879 | 0.009259 |
| 6.00 | 8.198 | 0.181 | 33.924 | 6.238 | −4.54 | 13.15 | 0.008514 | 0.008717 |

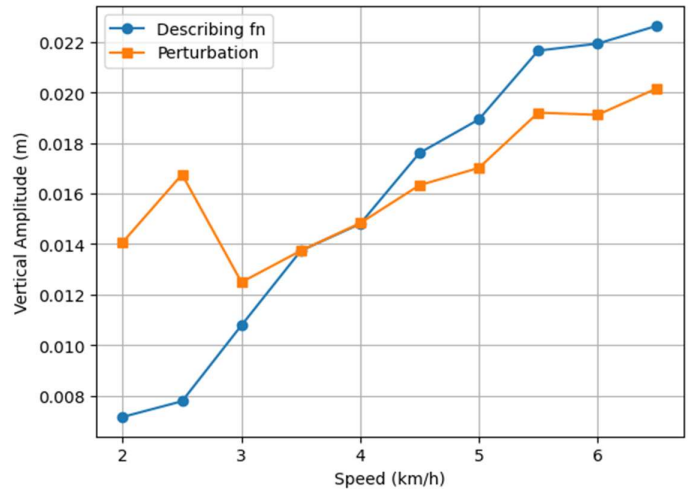

**Figure 3: First-order amplitude of vertical oscillator**

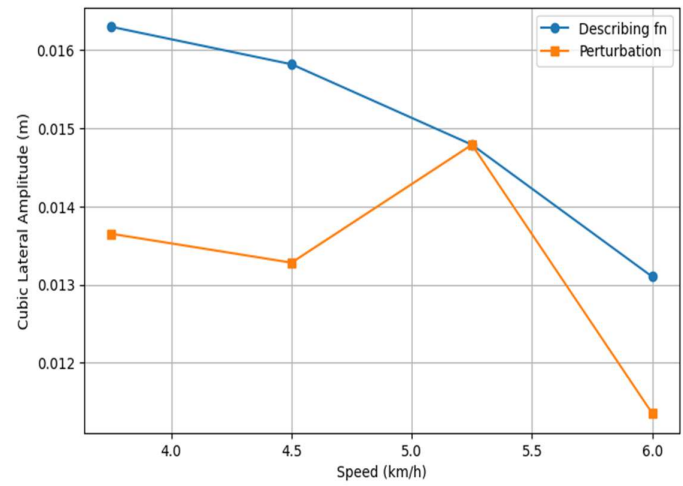

**Figure 4: First-order amplitude of cubic lateral oscillator**

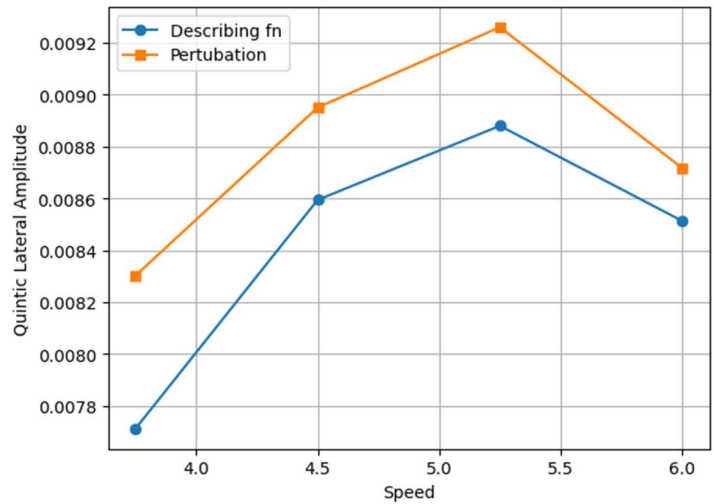

**Figure 5: First-order amplitude of quintic lateral oscillator**

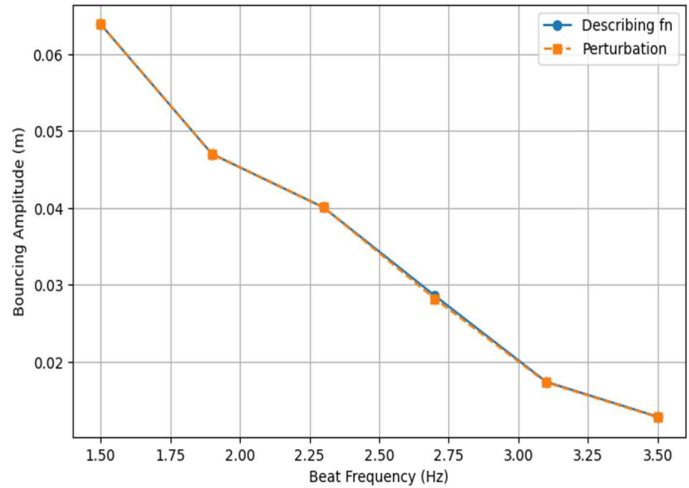

**Figure 6: First-order amplitude of bouncing oscillator**

7. **Future insights**

Interestingly, most of the natural processes are periodic [16,25–27], which has fuelled humans to model and explore the mathematical behaviour of these phenomena. Differential equations have predominantly been the mathematical tool to represent these dynamics [28]. In the mid-20th century, Norman Levinson [29] proposed a set of conditions that is sufficient to prove the existence of a periodic solution in a differential equation of the form $\ddot{x} + f(x, \dot{x})\dot{x} + g(x) = e(t)$. Researchers [19,30–35] have found less stringent sufficiency conditions to prove the existence of periodic solutions. In the past decade, new mathematical approaches have been developed in the literature [36–40] to prove the existence of periodic solutions in complicated nonlinear oscillators. An analytical proof of periodicity serves as a prerequisite canvas for the application of perturbation methods to estimate amplitudes and prove the stability of limit cycle [21,41–43]. According to Poincare-Bendixson theorem [18], if a phase path can be found that cannot escape from some bounded domain D of the phase plane, and if D contains no equilibrium points, then the theorem implies that the phase path must either be a closed loop or tend to a limit cycle. In either case, the system has a periodic solution. In the recent decade, several new methods of defining a bounded region to apply the annular Poincare- Bendixson theorem have been presented. [18,44,45]. Despite these advancements, an analytical proof for the existence of a periodic solution and a stable limit cycle has not been developed for these oscillators modelling human locomotion. Mathematical methods for functional analysis of non-standard and non-autonomous oscillators are other interesting domains to explore.

## 8. Conclusion

The article provides a mathematical framework for characterizing the key attributes of existing nonlinear self-sustained oscillators developed for modelling human walking and bouncing. This exercise is performed by focusing on demonstrations around the mathematically intricate vertical oscillator. The study also presents a concise overview of mathematical approaches to model human locomotion, with particular emphasis on the self-sustained oscillator framework, which forms the central theme of this work. The Krylov-Bogolyubov perturbation method is applied to show the stability of the limit cycle. The multiple-scales method is employed to examine the degree of asymptotic convergence, deducing an iterative procedure to estimate the order of timescale required to obtain an approximate solution up to a given order of magnitude. The relationship between various model parameters has been described using the describing function method, and an effective alternate amplitude equation for human bouncing has been obtained. This opens us up to a new metaverse of linking computational parameters to their physical significance and perhaps more efficient reasoning, replacing hit-and-trial or optimization-based estimation of parameters.

**Funding:** This research received no external funding.

**Conflicts of Interest:** The authors declare no conflict of interest.

**Declaration of Generative AI Use**

During the preparation of this work, Varun Nevash used a generative artificial intelligence tool to assist with language refinement and clarity in the presentation of the manuscript. After using this tool, the author carefully reviewed and edited the manuscript and takes full responsibility for the originality, accuracy, and integrity of the published work.

# References

[1] S. Živanović, A. Pavic, P. Reynolds, Vibration serviceability of footbridges under human-induced excitation: a literature review, Journal of Sound and Vibration 279 (2005) 1–74. https://doi.org/10.1016/j.jsv.2004.01.019.

[2] P. Dallard, T. Fitzpatrick, A. Flint, A. Low, R.R. Smith, M. Willford, M. Roche, London Millennium Bridge: Pedestrian-Induced Lateral Vibration, J. Bridge Eng. 6 (2001) 412–417. https://doi.org/10.1061/(ASCE)1084-0702(2001)6:6(412).

[3] T. Morbiato, R. Vitaliani, A. Saetta, Numerical analysis of a synchronization phenomenon: Pedestrian–structure interaction, Computers & Structures 89 (2011) 1649–1663. https://doi.org/10.1016/j.compstruc.2011.04.013.

[4] S. Erlicher, A. Trovato, P. Argoul, A modified hybrid Van der Pol/Rayleigh model for the lateral pedestrian force on a periodically moving floor, Mechanical Systems and Signal Processing 41 (2013) 485–501. https://doi.org/10.1016/j.ymssp.2013.08.005.

[5] S.K. Maurya, A. Kumar, P. Kumar, Modeling of Human-Induced Dynamic Bouncing Force Using a Self-Sustained Nonlinear Oscillator, Int. J. Str. Stab. Dyn. (2024) 2550178. https://doi.org/10.1142/S0219455425501780.

[6] V. Racic, A. Pavic, J.M.W. Brownjohn, Experimental identification and analytical modelling of human walking forces: Literature review, Journal of Sound and Vibration 326 (2009) 1–49. https://doi.org/10.1016/j.jsv.2009.04.020.

[7] F. Venuti, V. Racic, A. Corbetta, Modelling framework for dynamic interaction between multiple pedestrians and vertical vibrations of footbridges, Journal of Sound and Vibration 379 (2016) 245–263. https://doi.org/10.1016/j.jsv.2016.05.047.

[8] V. Racic, J. Chen, Data-driven generator of stochastic dynamic loading due to people bouncing, Computers & Structures 158 (2015) 240–250. https://doi.org/10.1016/j.compstruc.2015.04.013.

[9] M. Bocian, J.H.G. Macdonald, J.F. Burn, Biomechanically inspired modelling of pedestrian-induced forces on laterally oscillating structures, Journal of Sound and Vibration 331 (2012) 3914–3929. https://doi.org/10.1016/j.jsv.2012.03.023.

[10] B. Lin, S. Zhang, S. Živanović, Q. Zhang, F. Fan, Verification of damped bipedal inverted pendulum model against kinematic and kinetic data of human walking on rigid-level ground, Mechanical Systems and Signal Processing 200 (2023) 110561. https://doi.org/10.1016/j.ymssp.2023.110561.

[11] P. Kumar, A. Kumar, S. Erlicher, A modified hybrid Van der Pol–Duffing–Rayleigh oscillator for modelling the lateral walking force on a rigid floor, Physica D: Nonlinear Phenomena 358 (2017) 1–14. https://doi.org/10.1016/j.physd.2017.07.008.

[12] P. Kumar, A. Kumar, S. Erlicher, A Nonlinear Oscillator Model to Generate Lateral Walking Force on a Rigid Flat Surface, Int. J. Str. Stab. Dyn. 18 (2018) 1850020. https://doi.org/10.1142/S0219455418500207.

[13] P. Kumar, A. Kumar, V. Racic, Modeling of Longitudinal Human Walking Force Using Self-Sustained Oscillator, Int. J. Str. Stab. Dyn. 18 (2018) 1850080. https://doi.org/10.1142/S0219455418500803.

[14] P. Kumar, A. Kumar, V. Racic, S. Erlicher, Modelling vertical human walking forces using self-sustained oscillator, Mechanical Systems and Signal Processing 99 (2018) 345–363. https://doi.org/10.1016/j.ymssp.2017.06.014.

[15] S. Erlicher, A. Trovato, P. Argoul, Modeling the lateral pedestrian force on a rigid floor by a self-sustained oscillator, Mechanical Systems and Signal Processing 24 (2010) 1579–1604. https://doi.org/10.1016/j.ymssp.2009.11.006.

[16] H.G. Mayr, J.-H. Yee, M. Mayr, R. Schnetzler, Nature's autonomous oscillators, NS 04 (2012) 233–244. https://doi.org/10.4236/ns.2012.44034.

[17] Pikowski, A., Rosenblum, M., Kurths, J. (2001) Synchronization: A Universal Concept in Nonlinear Science. 1st edn. (Cambridge University Press, Cambridge, UK), n.d.
[18] R.D. Gregory, Classical mechanics: an undergraduate text, Repr, Cambridge Univ. Press, Cambridge, 2007.
[19] Dominic Jordan and Peter Smith, Nonlinear Ordinary Differential Equations An Introduction for Scientists and Engineers Fourth Edition, n.d.
[20] Rick Sengers, Multiple scales and related methods applied to ordinary differential equations, (n.d.).
[21] S. Saha, G. Gangopadhyay, The existence of a stable limit cycle in the Liénard–Levinson–Smith (LLS) equation beyond the LLS theorem, Communications in Nonlinear Science and Numerical Simulation 109 (2022) 106311. https://doi.org/10.1016/j.cnsns.2022.106311.
[22] Z. Úředníček, Nonlinear systems - describing functions analysis and using, MATEC Web Conf. 210 (2018) 02021. https://doi.org/10.1051/matecconf/201821002021.
[23] E. Vidal, A. Poveda, M. Ismail, Describing functions and oscillators, IEEE Circuits Devices Mag. 17 (2001) 7–11. https://doi.org/10.1109/101.968910.
[24] D.G. Robertson, J.H. Lee, On the use of constraints in least squares estimation and control, Automatica 38 (2002) 1113–1123. https://doi.org/10.1016/S0005-1098(02)00029-8.
[25] P.H. Richter, J. Ross, Oscillations and efficiency in glycolysis, Biophysical Chemistry 12 (1980) 285–297. https://doi.org/10.1016/0301-4622(80)80006-8.
[26] P.E. Rapp, Why are so many biological systems periodic?, Progress in Neurobiology 29 (1987) 261–273. https://doi.org/10.1016/0301-0082(87)90023-2.
[27] van der Pol, B., van der Mark, J. The heartbeat considered as a relaxation oscillation and an electrical model of the heart (1928) Philosophical Magazine Supplement, 6, pp. 763-775., (n.d.).
[28] V.N. Krishnachandran, Differential Equations: A Historical Refresher, (2020). https://doi.org/10.48550/ARXIV.2012.06938.
[29] N. Levinson, On The Existence of Periodic Solutions For Second Order Differential Equations with a Forcing Term, Journal of Mathematics and Physics 22 (1943) 41–48. https://doi.org/10.1002/sapm194322141.
[30] G. Villari, Extension of some results on forced nonlinear oscillations, Annali Di Matematica Pura Ed Applicata 137 (1984) 371–393. https://doi.org/10.1007/BF01789402.
[31] J. Zu, Existence and Uniqueness of Periodic Solution for Nonlinear Second-Order Ordinary Differential Equations, Boundary Value Problems 2011 (2011) 1–11. https://doi.org/10.1155/2011/192156.
[32] I. Muntean, Harmonic oscillations for some systems of two differential equations, Journal of Mathematical Analysis and Applications 24 (1968) 474–485. https://doi.org/10.1016/0022-247X(68)90003-6.
[33] C.E. Langenhop, Note on Levinson's Existence Theorem for Forced Periodic Solutions of a Second Order Differential Equation, Journal of Mathematics and Physics 30 (1951) 36–39. https://doi.org/10.1002/sapm195130136.
[34] T.A. Burton, C.G. Townsend, On the generalized Liénard equation with forcing function, Journal of Differential Equations 4 (1968) 620–633. https://doi.org/10.1016/0022-0396(68)90012-0.
[35] K.W. Chang, The Existence of Forced Periodic Solutions for Second Order Differential Equations, Journal of Mathematics and Physics 46 (1967) 93–98. https://doi.org/10.1002/sapm196746193.

[36] M. Bayat, B. Mehri, A necessary condition for the existence of periodic solutions of certain three dimensional autonomous systems, Applied Mathematics Letters 22 (2009) 1292–1296. https://doi.org/10.1016/j.aml.2009.01.045.

[37] B. Mehri, M.A. Niksirat, On the existence of periodic solutions for certain differential equations, Journal of Computational and Applied Mathematics 174 (2005) 239–249. https://doi.org/10.1016/j.cam.2004.04.011.

[38] B. Mehri, M. Niksirat, On the existence of periodic solutions for the quasi-linear third order system of O.D.Es, Proc. Appl. Math. Mech. 1 (2002) 520. https://doi.org/10.1002/1617-7061(200203)1:1<520::AID-PAMM520>3.0.CO;2-R.

[39] B. Mehri, D. Shadman, On the existence of periodic solutions of a certain class of second order nonlinear differential equation, Mathematical Inequalities & Applications (1998) 431–436. https://doi.org/10.7153/mia-01-41.

[40] R.N. Jazar, M. Mahinfalah, B. Mehri, Third-order systems: Periodicity conditions, International Journal of Non-Linear Mechanics 44 (2009) 855–861. https://doi.org/10.1016/j.ijnonlinmec.2009.05.006.

[41] M.Z. Alam, Md. Alal Hosen, M.S. Alam, A new modified Lindstedt–Poincare method for nonlinear damped forced oscillations, Results in Physics 51 (2023) 106673. https://doi.org/10.1016/j.rinp.2023.106673.

[42] C.-S. Liu, Y.-W. Chen, A Simplified Lindstedt-Poincaré Method for Saving Computational Cost to Determine Higher Order Nonlinear Free Vibrations, Mathematics 9 (2021) 3070. https://doi.org/10.3390/math9233070.

[43] Nayfeh, A.H. (1985) Problems in Perturbation. Wiley, New York, USA, n.d.

[44] H. Chen, Y. Tang, D. Xiao, Global dynamics of hybrid van der Pol–Rayleigh oscillators, Physica D: Nonlinear Phenomena 428 (2021) 133021. https://doi.org/10.1016/j.physd.2021.133021.

[45] H. Chen, L. Zou, Global study of Rayleigh–Duffing oscillators, J. Phys. A: Math. Theor. 49 (2016) 165202. https://doi.org/10.1088/1751-8113/49/16/165202.